# "Corotating Interaction Regions (CIRs)", "Interaction Regions (IRs)" and "Stream Interaction Regions (SIRs)", which term should be used?

**Bruce T. Tsurutani[1*], Rajkumar Hajra[2], and Gurbax S. Lakhina[3]**

[1] Retired, Pasadena, California 91208, USA
[2] School of Earth and Space Sciences, University of Science and Technology of China, Hefei 230026, China
[3] Retired, B-701, Neel Sidhi Towers, Plot no 195, Sector-12, Vashi, Navi Mumbai 400703, India
[*] *Correspondence to*: Bruce T. Tsurutani (bruce.tsurutani@gmail.com)

**Abstract.** We discuss the early history of quasiperiodic ~27-day recurrent geomagnetic activity starting with Maunder (1904, 1905), Chree (1913) and Bartels (1932, 1934), and the Bartels term "M-regions". We show the iconic "interaction region (IR)" schematic of Belcher and Davis (1971) and the further development of Smith and Wolfe (1976) and the term "corotating interaction region (CIR)". We quote the Jian et al. (2006) definition of a "stream interaction region (SIR)". We disagree with Jian et al. (2006) on the use of the term (SIR) to indicate "transient and possibly localized stream interactions" with "poor recurrence" (Gosling et al., 2001). We feel that this description is too vague for use in scientific studies. We suggest, instead identifying the specific known interplanetary phenomena: interplanetary coronal mass ejection (ICME) sheaths, ICMEs (loops, magnetic clouds, filaments), CIRs, high-speed streams (HSSs) and slow streams. All of these various interplanetary phenomena have different solar and interplanetary origins and different plasma and magnetic field properties. The different interplanetary phenomena have been shown to have different geomagnetic effectivenesses. In keeping with this theme of naming specific interplanetary phenomenon, we introduce the term "Super CIR (SCIR)", which describes a CIR associated with magnetic reconnection at the edge of a solar coronal hole with an embedded coronal jet. SCIRs are a new form of a "transient event" and can be identified by exceptionally strong internal magnetic fields and bounded by both forward and reverse shocks. The SCIR on 6–7 April 2000 caused an exceptionally strong SYM-H = –319 nT superstorm, a first detected/reported event of its kind.

## 1 History

Maunder (1904) was perhaps the first person to report the statistics of “magnetic disturbances of the solar rotation-period” (the phrase is taken from the title of the Maunder, 1905 paper). In the 1904 paper, Maunder used the Greenwich, England observations of “magnetic movements” for his study. In the study of data from 1886 to 1889 and 1895 to 1898, he found intervals where “there is a strong tendency for certain (solar) longitudes to recur”. Maunder (1904) concluded: “Several important consequences follow from this relation. First, that our magnetic disturbances are directly due to some solar action. Next, that action must be located in certain restricted areas of the Sun’s surface; it cannot be general to the surface as a whole, for it is precisely as certain meridians return to the centre of the disc, that we have the return of the disturbances. Thirdly, the mode of the transmission of this solar influence to the Earth, must be along definite lines; it cannot be of the nature of radiation, equal in all directions, as it is with light and heat. These are the most important conclusions to be drawn from the inspection of the catalogue.”

It was not until Chree (1913), the Director of The King’s Observatory, Richmond, U.K., proved that Maunder’s ~27-day quasiperiodic results were statistically significant, giving us the “Chree superposed epoch” statistical analyses, a method widely in use today. Although these periodic geomagnetic activities at Earth were substantiated by Chree, no identifiable optical features causing them were apparent on the Sun’s visible disc.

While analyzing the data of terrestrial-magnetic activity for the years 1906–1931, Bartels (1932) similarly found strong ~27-day recurrences related to solar rotation. Because he could not identify any visible signatures on the Sun that could cause ~27-day periodic geomagnetic activity at Earth, Bartels (1932, 1934) called these “magnetically active” or “M-regions” of the Sun. After 1934, the science community adopted the Bartels M-region terminology/name.

It was not until soft X-ray images of the Sun were available from the Skylab satellite that Krieger et al. (1973) “identified a magnetically open structure in the low corona……with density scale heights typically a factor of two less than that in the surrounding large scale magnetically closed region”. Krieger et al. (1973) traced back a high velocity stream using “instantaneous ideal spirals” and found “a striking agreement between the Carrington longitude of the solar source of a recurrent high velocity solar wind stream with the position of the hole”. Krieger et al. (1973) called these magnetically open structures “coronal holes”.

However, it was postulated by later scientists that the coronal hole high-speed solar winds would not simply propagate unimpededly from the Sun to the Earth, but would interact with an upstream slow solar wind. Belcher and Davis (1971) showed this schematically as in Figure 1, and called this an "interaction region" or IR.

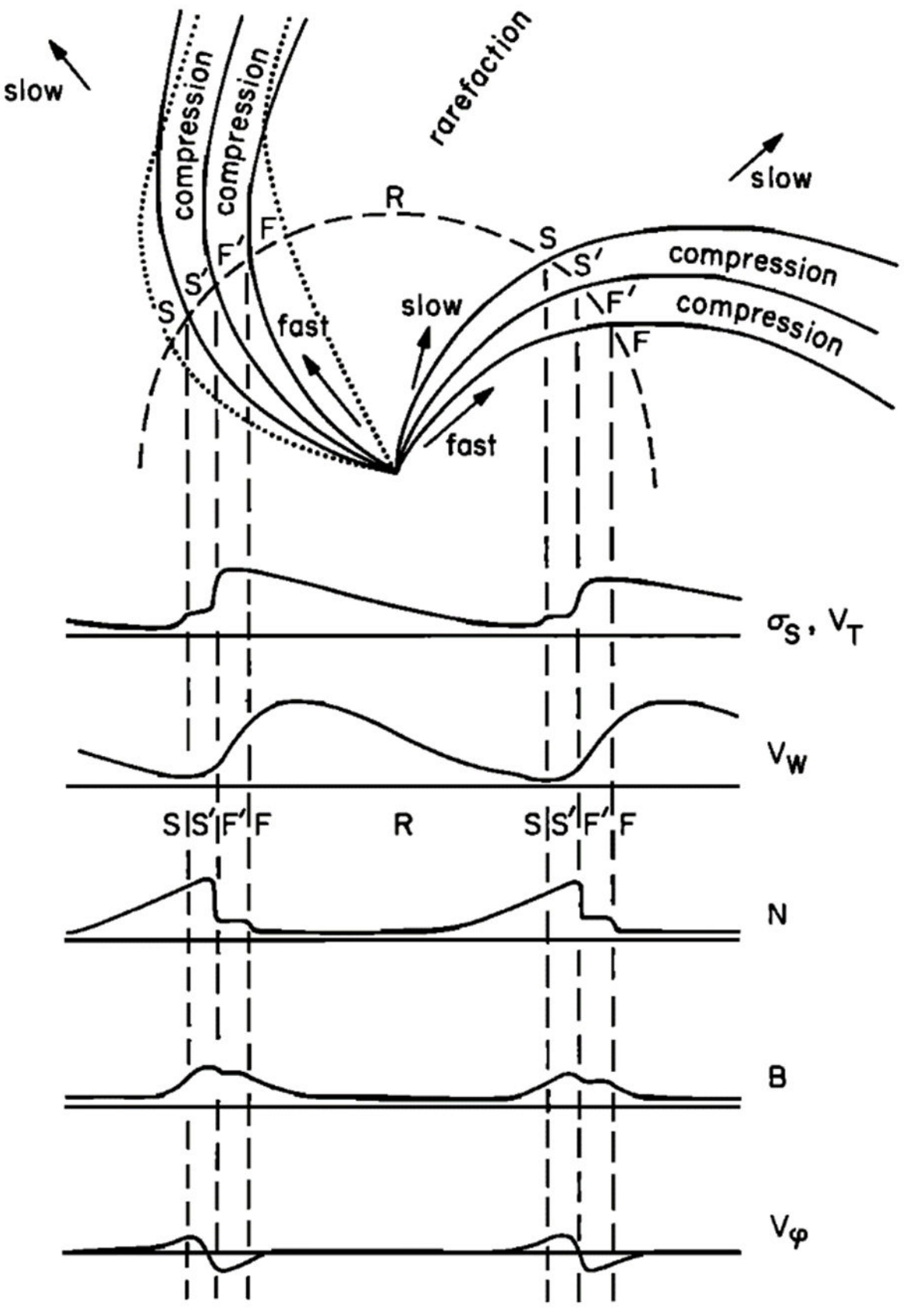

**Figure 1.** Top: schematic of two high-speed streams and adjacent slow stream corotating with the Sun. The regions indicated are: the unperturbed slow solar wind (S), compressed, accelerated slow solar wind ($S^{/}$), compressed, decelerated fast solar wind ($F^{/}$), unperturbed fast solar wind (F), and a rarefaction (R). $S^{/}$ and $F^{/}$ form the interaction region, and the stream interface is at the $S^{/}$–$F^{/}$ boundary. Dotted lines indicate magnetic field lines in the slow and fast solar wind which thread into the interaction region beyond 1 AU. Bottom: curves showing as functions of time the changes in solar wind parameters that will be observed by a spacecraft as the streaming pattern sweeps past at 1 AU. Various plasma parameters: proton thermal speed ($V_T$), magnetic field fluctuation level ($\sigma_s$), solar wind speed ($V_W$), density (N), magnetic field intensity (B), and transverse component of the solar wind velocity ($V_\varphi$). The figure is taken from Belcher and Davis (1971).

Belcher and Davis (1971) postulated that the "interaction region" was not corotating in a physical sense, but was shaped like an Archimedean spiral. They stated: "It is instructive to consider this steady state flow in a rotating coordinate system. The structure in the upper half of their Fig. 13 (Figure 1 in this article) now does not rotate; instead, the spacecraft moves clockwise in a circle and makes observations that, when plotted as functions of time, yield the idealized curves shown in the bottom half of the figure. In this corotating frame, the velocity is everywhere parallel to the smoothed magnetic field lines, and hence the flow is in a spiral whose pitch changes as it passes into the regions of compression because of the pressure gradient (or discontinuity) across the transition. The deflection provides a natural explanation for the observation."

Belcher and Davis (1971) were later challenged by Burlaga (1974) and Hundhausen and Burlaga (1975) concerning a sharp transition between slow and fast flows (the formation of a tangential discontinuity as stated by Belcher and Davis, 1971) separating the two flows and plasma and magnetic field regions (see Richardson, 2018 for further discussion). However, the tangential discontinuity was later found in the plasma data and is now referred to as a "stream interface" or SI (Gosling et al., 1978).

Smith and Wolfe (1976) performed the first high spatial resolution examination of these "interaction regions" using Pioneer 10 and 11 magnetometer and plasma data. They called these regions "Corotating Interaction Regions" or CIRs due to the shape of the CIRs in space. The shape is like water coming from a rotating water sprinkler. The water comes radial outward (with a small aberrational effect), but because of the rotation of the sprinkler head, the shape of the water has a "corotational shape". Smith and Wolfe (1976) stated: "As the solar wind expands beyond 1 AU, this stream interaction becomes more pronounced (Collard and Wolfe, 1974). Compression of the plasma and magnetic field within the interaction region leads to stresses which accelerate the preceding slow plasma and decelerate the trailing fast plasma. According to theory, these interaction regions trace out a spiral which corotates with the sun (Siscoe, 1972)."

Smith and Wolfe (1976) showed the presence of shocks at the leading and trailing edges of the CIRs at large distances from the Sun and the general lack of shocks closer to the Earth, advancing the knowledge of Belcher and Davis (1971) on this topic. An example of a shock pair associated with the slow and fast stream interaction region is shown schematically in Figure 2.

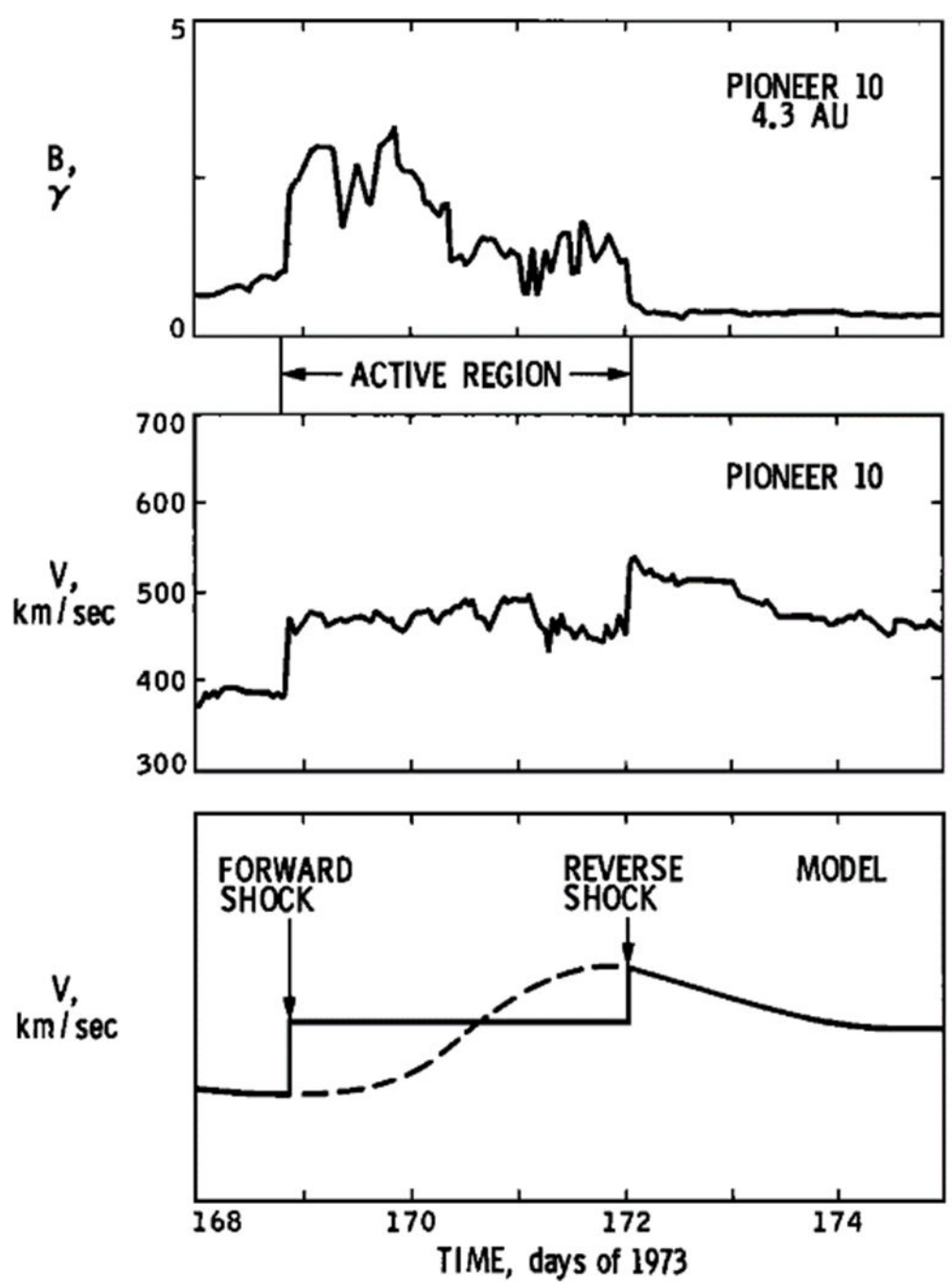

**Figure 2.** Pioneer 10 plasma and magnetic field observations. Top panel: hourly averages of the field magnitude. The space between an abrupt onset and an equally abrupt termination defines the "interaction region". Central panel: hourly values of the solar wind velocity with abrupt jumps at the beginning and end of the "interaction region". Bottom panel: a qualitative diagram showing how the positive gradient associated with a fast stream at 1 AU (dashed) is replaced at large distances by a region of essentially zero gradient bounded by a forward and a reverse shock. The figure is taken from Smith and Wolfe (1976).

Tsurutani et al. (1995b) and Tsurutani et al. (2006) demonstrated that it was amplification of interplanetary Alfvén waves within the two compressed regions of the CIR that caused enhanced geomagnetic activity at the Earth. Intense, compressed interplanetary magnetic field (IMF) $B_z$ southward components of the Alfvén waves within the CIRs, through magnetic reconnection with the Earth's magnetopause magnetic fields (Dungey, 1961; Tsurutani and Meng, 1972), could cause geomagnetic storms (SYM-H $< -50$ nT). However, because of the high level of IMF $B_z$ fluctuations (fluctuating in both the northward and southward directions) found within the CIRs (Tsurutani et al., 1995a), the storm intensities rarely exceed SYM-H $< -100$ nT (Tsurutani et al., 2024). It was also noted that CIRs would sometimes contain primarily IMF $B_z$ northward components and thus, little or no geomagnetic activity would result. These latter phenomena explain the statistical nature of the Maunder (1904, 1905) and Bartels (1932) results. Sometimes there would be geomagnetic activity in the next 27 days and sometimes not. Besides the CIR there is a trailing high-speed solar wind stream which can impact the Earth's magnetosphere for days to weeks, depending on the size of the coronal hole (Tsurutani and Gonzalez, 1987; Tsurutani et al., 1995a; Hajra et al., 2013). The pure high-speed solar wind stream following the CIR contains nonlinear Alfvén waves which could cause long durations of low-level geomagnetic activity. The authors explained that this geomagnetic activity was not a CIR storm "recovery phase", but was fresh solar wind energy input into the magnetosphere through magnetic reconnection. Tsurutani and Gonzalez (1987) named these geomagnetically active intervals "High Intensity, Long-Duration, Continuous AE Activity" or HILDCAA events. See also Hajra et al. (2013, 2014a, c, b, 2015a, b).

Between 1976 and 2006 the "interaction regions" of Belcher and Davis (1971) were called CIRs in the literature following Smith and Wolfe (1976). However, in 2006, Jian et al. (2006) called the CIRs inside 1 AU, "stream interaction regions" or SIRs. Jian et al. (2006) wrote: "A stream interaction region (SIR) forms when a fast solar stream overtakes a slow stream, leading to structure that evolves as an SIR moves away from the Sun." They conducted "a separate assessment of the longer-lasting corotating interaction regions (CIRs) that recur on more than one solar rotation." Jian et al. (2006) thus made a distinction between when an "interaction region" recurred 27 days later from those that did not.

Jian et al. (2006) may have overlooked the point made in the Smith and Wolfe (1976) and Belcher and Davis (1971) papers that a CIR/IR showed the dynamics of the interplanetary interaction between a fast solar wind stream and a slow solar wind stream and not the recurrence of fast stream after 27 days.

Jian et al. (2006) also mentioned that they used the term "stream interaction regions (SIRs)" "following the suggestion of Gosling et al. (2001)" to include "transient and possibly localized stream interactions" "with poor recurrence." We find this categorization extremely vague and confusing as to what are these "transient stream interactions" and how would one distinguish them from observations of regular interaction regions?

Many of these "transient and possibly localized stream interactions" have been carefully identified in the literature, and have been used to relate to geomagnetic activity. Sheaths upstream of interplanetary coronal mass ejections (ICMEs) have been identified by Tsurutani et al. (1988). Sheaths, which are plasma and magnetic field regions, which originate from the slow solar wind, have been noted to cause extremely intense magnetic storms (e.g. Meng et al., 2019). Coronal mass ejections (CMEs) near the Sun have been shown to be composed of three essential parts: outer loops, a dark region and a filament (Illing and Hundhausen, 1986). These various parts have been detected at 1 AU. A loop was detected by Tsurutani et al. (1998), magnetic clouds (MCs, the dark region near the Sun) by Burlaga et al. (1981), and filaments by Burlaga et al. (1998). A listing of many filaments detected in interplanetary space has been catalogued by Lepri and Zurbuchen (2010). Zhang et al. (2007) and Echer et al. (2008) have identified the interplanetary causes of Dst ≤ –100 nT magnetic storms. They found that the majority of these major magnetic storms were caused by either the MC portion of ICMEs or their upstream sheaths. CIRs were rarely the cause of Dst ≤ –100 nT magnetic storms (Tsurutani et al., 1995a; Echer et al., 2008). We suggest readers use Zhang et al. (2007) and Echer et al. (2008) and their examples to help them identify specific interplanetary phenomena if they wish to determine the causes of geomagnetic activity for any events under their study.

Richardson (2018) has written a review of "stream interaction regions" throughout the heliosphere. In the abstract, he wrote: "This paper focuses on the interactions between the fast solar wind from coronal holes and the intervening slower solar wind, leading to the creation of stream interaction regions that corotate with the Sun and may persist for many solar rotations." The above description of an SIR is the same as described by Belcher and Davis (1971) and Smith and Wolfe (1976) for a CIR. More recently Chi et al. (2018) and Nguyen et al. (2025) have followed Richardson (2018) in calling all CIRs as SIRs, complicating things further.

Besides the lack of a need for a new term like SIR (to describe the same CIR events), we suggest that the term is causing unnecessary confusion. As previously mentioned, the term SI has been used to indicate the stream interface or tangential discontinuity between the compressed, accelerated slow solar wind and the compressed, decelerated high-speed solar wind. People are confusing the term SIR to mean Stream Interface Region.

Based on the above discussion, we introduce a new type of ejecta in the solar wind. In the original article (Tsurutani et al., 2024) it was called a CIR. However, because it is thought to be a highly unusual CIR with a different solar origin (we will discuss its properties below), it deserves a new name. We will call this a Super CIR or SCIR. The event on 6–7 April 2000 shown in Figure 3 was originally misidentified by several authors (e.g. Huttunen et al., 2002; Meng et al., 2019) as an ICME and was recently reanalyzed (Tsurutani et al., 2024) as a highly unusual CIR. The interplanetary structure was unusual in that it was bounded by both a very strong fast forward shock (FS; magnetosonic Mach number ~4.6) and a fast reverse shock (RS; magnetosonic Mach number ~1.9). The peak magnetic field magnitude was an unusually high ~33 nT with an unusually high peak $B_z$ of –27 nT. These field strengths are ~2 to 3 times normal CIR field strengths. It is speculated that the high magnetic field region is an embedded coronal jet associated with magnetic reconnection, which occurred at the edge of a contracting coronal hole. This event is believed to have a very different solar source than typical CIRs. As an aside, this SCIR caused an unusually strong magnetic storm of peak SYM-H intensity = –319 nT. However, the resultant geomagnetic activity should not be considered a criterion for identifying a SCIR. In the future, SCIRs will be discovered, which have northward directed $B_z$ magnetic fields with geomagnetic quiet resulting.

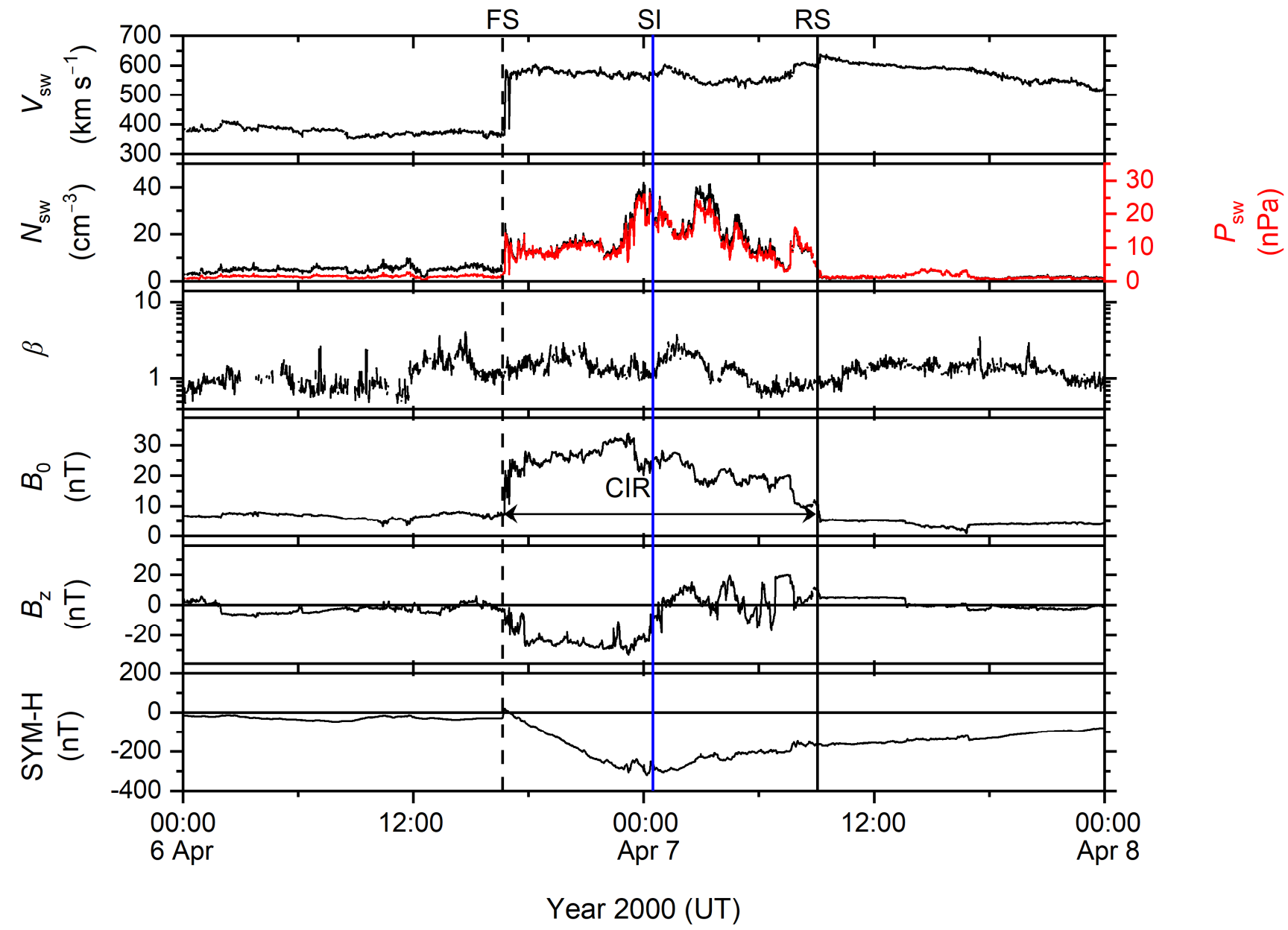

**Figure 3.** An SCIR detected at 1 AU. From top to bottom, panels show variations of the solar wind plasma speed $V_{sw}$, density $N_{sw}$ (black, legend on the left) and ram pressure $P_{sw}$ (red, legend on the right) in the same panel, plasma-$\beta$, IMF magnitude $B_0$, $B_z$-component, and the ring current index SYM-H during 6–7 April 2000. Vertical lines indicate a fast forward shock (FS, black dashed line), a stream interface (SI, blue solid line), and a reverse shock (RS, black solid line). The CIR, bounded by the fast forward and reverse shocks, caused a magnetic storm of peak SYM-H intensity = –319 nT. The figure is updated from Tsurutani et al. (2024).

## 2 Discussion

The term “transient and possibly local stream interaction regions” used by Gosling et al. (2001) and Jian et al. (2006) is vague and not useful. Specific terms such as ICME sheaths, solar wind high-speed streams (HSSs), slow solar wind streams, ICMEs (with subcategories of loops, MCs, and filaments), CIRs and SCIRs should be used. We direct the readership to Zhang et al. (2007) and Echer et al. (2008) and their examples of sheaths, MCs and CIRs, Burlaga et al. (1998) and Lepri and Zurbuchen (2010) for examples of filaments, and Tsurutani et al. (2024) for an example of an SCIR. Any of these authors or coauthors would be willing to help people new to the field to identify specific interplanetary phenomena.

We find the term SIR does not indicate anything different from the Belcher and Davis (1971) and Smith and Wolfe (1976) IR/CIR definition concerning the physical nature of the "interaction region" close to the Sun, at 1 AU or distances beyond 1 AU. The name IR/CIR was given for the shape of the phenomenon in interplanetary space, like water issuing from a rotating lawn sprinkler head. We recommend that the name CIR be continued for use in the future.

*Data availability*. Figure 1 is taken from Belcher and Davis (1971). Figure 2 is taken from Smith and Wolfe (1976). Figure 3 was taken from Tsurutani et al. (2024).

*Author contributions*. BTT had the original idea and prepared the first draft. RH and GSL participated in the development, revision of the article and approved the final draft.

*Competing interests*. The author declares that there is no conflict of interest.

*Financial support*. The work of RH is funded by the "Research Fund for International Excellent Young Scientists" of the National Natural Science Foundation of China (NSFC) grant No. W2532030, the "Excellent Young Scientists Fund Program (Overseas)" of NSFC, and the "Hundred Talents Program" of the Chinese Academy of Sciences (CAS).